\documentclass[12pt]{article}

\usepackage{amssymb}

\def\fnote#1#2{\begingroup\def\thefootnote{#1}\footnote{#2}\addtocounter{footnote}{-1}\endgroup}

\begin{document}

\hfill {\bf CERN$-$PH$-$TH/2012$-$104}


\baselineskip=20pt
\parindent=0pt

\vskip .9truein

\centerline{\Large {\bf Emergent Spacetime and Black Hole Probes }} 
 \vskip .1truein
\centerline{\Large {\bf from Automorphic Forms}}

\vskip .3truein

\centerline{{\sc~ Rolf Schimmrigk$^{1,2}$}\fnote{}{$^1$On leave from Indiana University South Bend}
  \fnote{}{$^2$netahu@yahoo.com, rschimmr@iusb.edu}}

\vskip .2truein

\centerline{Theory Division, CERN}
\centerline{1211 Geneva 23, Switzerland~}

\vskip 1truein

\centerline{\bf Abstract}
\begin{quote}
 Over the past few years the arithmetic Langlands program has found
 applications in two quite different problems that arise in string physics.
 The first of these is concerned with the fundamental problem of deriving
 the geometry of spacetime from the worldsheet dynamics, leading to a
 realization of the notion of an emergent spacetime in string theory.
 The second problem is concerned with the idea of using automorphic black
 holes as probes of spacetime. In this 
 article both of these applications of 
the Langlands program are described.
\end{quote}

\renewcommand\thepage{}
\newpage
\parindent=0pt

 \pagenumbering{arabic}

\baselineskip=20pt
\parskip=.04truein

\tableofcontents

\vskip .5truein

\baselineskip=20.6pt
\parskip=.15truein

\section{Introduction}\label{ra_sec1}

Applications of arithmetic algebraic geometry have been rare in
particle physics in general and are not common in string theory. One reason for this
might be that the methods involved are far removed from the tools
contained in the standard kit issued to theoretical physicists.
There is one part of number theory, however, that one might expect
to be of relevance for string physics, and that is the theory of
modular forms, in particular the link of modular forms to
geometry. This is an old subject, with roots that can be traced
back to the work of Klein and his student
Hurwitz more than a century ago. This line of thought was continued, 
with interruptions, by Hecke, Eichler, Taniyama, Shimura, Weil, as well as 
Wiles and Taylor, in the context of elliptic curves. Its relevance for string
physics, however, depends in an essential way on the important
extension of the elliptic setting by Grothendieck, Langlands, Deligne,
Serre, and others, into what today is called the Langlands
program.

The framework provided by the Langlands program consists of
several difficult conjectures about automorphic forms, of which
the most important for the applications discussed here is what is
often called the reciprocity conjecture. It is concerned with the
geometric origin of automorphic forms and their associated
representations and in very rough terms states that all motives
are automorphic. In a first approximation motives can be thought
of as geometric building blocks, whose lego-like structure allows
to build full-fledged manifolds. While the inverse of this
conjecture is not believed to be true, at least within the current
framework of both automorphic forms and motives, it is expected
that a certain sub-class of automorphic forms is motivic.

The two ideas described in this review are both concerned with the
geometric nature of automorphic forms that arise in string theory.
The first addresses the old problem of deriving the geometry of
spacetime from string physics. Given that string theory is
expected to be a complete theory one might expect that it should
be possible to derive not only a model for particle physics that
extends the standard model, but also the arena where particle 
 interactions
takes place, i.e. the structure of spacetime itself. This vague
expectation can be made precise within the framework of the
Langlands program because the conformal field theory on the string
worldsheet leads to modular forms that encode the fundamental
physics of the string. One can therefore ask whether these modular
forms can be used to construct geometric motives that provide the
building blocks of the extra dimensions. By combining several of
these motives one can hope to build the compactification manifold
from scratch. The review of this problem is based mostly on the
work of \cite{rs08,kls10}, and references therein.

The second problem is concerned with a new application of black
holes that can be suggested in light of the Langlands program. In
this context the automorphic forms arise as functions that
describe the entropy of black holes in certain classes of string
compactifications with extended supersymmetry. Given such
automorphic black hole entropy functions, one can ask whether they
encode motivic information about the compact part of spacetime,
leading to the view that automorphic black holes can be used as
physical probes that are sensitive to the nature of the extra
dimensions. The modularity of the entropy implies that if one were
to experiment with such black holes in the laboratory, a finite
number of measurements would suffice to completely determine these
functions. The review of this problem is based on \cite{cs11}.

\section{Emergent spacetime from string modular forms}

Speculations about the nature of spacetime have a long history
which can be traced back at least to Born's 1919
pre-Heisenberg thinking that a true understanding of quantum
theory will very likely necessitate a drastic change in the view
of spacetime as a manifold over the real numbers \cite{g05}. In
more recent times this has led to the speculation that spacetime
might be $p-$adic in nature. A more natural view would be to
combine all $p-$adic numbers into an adelic structure.

The view that at some fundamental level spacetime is adelic in
nature clashes with the fact that in physics the dynamics is 
encoded in differential equations, which motivates the assumption
that spacetime is defined over the field of real numbers ${\mathbb
R}$, in order to be able to do analysis. In the case of higher
dimensional theories with extra dimensions supersymmetry leads to
compact spaces that are defined over the complex numbers ${\mathbb
C}$.  A more pragmatic point of view of number theoretic methods
in this complex framework is to think of restrictions of spacetime
to number fields, instead of ${\mathbb R}$ or ${\mathbb C}$, as a
particular type of lattice approximation, not necessarily as a radical
reinterpretation of the nature of spacetime. It is this more
pragmatic latter attitude that has led to some results in the
understanding of how space could be constructed from the physics
on the worldsheet. This does not preclude that adelic features
might eventually emerge as fundamental properties, but such an
assumption is not necessary for physical applications of number
theoretic methods.

\subsection{Emergent space from worldsheet modular forms}

The basic idea of the emergent spacetime program
initiated in \cite{rs01,su02}, and further developed in \cite{rs06,rs08}, 
is to use modular forms that
arise in the conformal field theoretic models on the worldsheet
$\Sigma$ to derive the geometric structure of the extra compact
dimensions in the spacetime manifold $X$. This program can be
viewed as a string theoretic refinement of the Langlands program,
part of which leads to the expectation that a particular class of
automorphic forms are supported by geometric structures called
motives, denoted here generically by $M$ \cite{g60s,l78, l79}. Motives
in turn can be viewed as support structures for certain types of
cohomology groups. While these motivic cohomology groups are not
the ones that have traditionally been encountered in physics, there exist certain
compatibility theorems which imply that their rank is identical to
that of cohomology groups that are more familiar from string
theory, such as the Hodge decomposition of the de Rham cohomology
for complex varieties.

The number theoretic nature of these motives can not be avoided
because the only known way to obtain automorphic forms from these
objects $M$ is via the computation of their $L-$function $L(M,s)$,
a function in the complex variable $s\in {\mathbb C}$ that a
priori is only defined over part of the complex plane ${\mathbb
C}$. It is, however, conjectured that this function can be
continued to all $s$, and that it satisfies a functional equation.
This functional equation encodes Poincar\'e duality of the
motives, and is a first indication of the automorphy of the
associated $q-$series.

The results obtained so far are mostly concerned with modular motives in the
sense that the inverse Mellin transform $f(M,s)$ of the motivic
$L-$series $L(M,s)$ leads to modular forms, in particular to cusp
forms of weight $w$ and level $N$ with respect to the Hecke
congruence subgroup
 \begin{equation}
 f(M,q) ~\in ~ S_w(\Gamma_0(N)).
 \end{equation}
 Here the weight $w$ of the form is determined by the weight ${\rm wt}(M)$ of
 the motive $M$ via
 \begin{equation}
 w(f) ~=~ {\rm wt}(M) ~+~ 1,
 \label{weight}\end{equation}
 where the weight wt($M$) of the motive can be viewed as the
 degree of the associated cohomology group
 \begin{equation}
   {\rm wt}(M) ~:=~ {\rm deg}~H(M) ~=~ r
 \end{equation}
 for $H(M) \subset H^r(X)$. This relation assumes that the motive 
 $M$ is pure, i.e. it arises within a smooth variety. This is the
 situation that applies to the case of weighted Fermat
 hypersurfaces considered in \cite{rs06,rs08,kls08}. The case of
 mixed motives, much less understood, applies in the context of
 phase transitions between Calabi-Yau varieties and has been
 considered in \cite{kls10} in the context of extending this program from the 
case of weighted Fermat spaces to families of manifolds.

\subsection{Construction of motives}

One of the problems encountered in the context of finding a string
theoretic interpretation of motivic automorphic forms is that
 in general a Calabi-Yau variety of arbitrary dimension has
 several nontrivial cohomology groups $H^r(X)$. For each
 of these groups a decomposition into their irreducible
 motivic subgroups
 \begin{equation}
 H^r(X) ~= ~ \bigoplus_i H(M_i)
 \end{equation}
 does not appear to be known in mathematics. Given a variety, the
 first question thus becomes how one should construct motives.
 For the simplest Calabi-Yau spaces, i.e. elliptic curves $E$, this is
 clear because there is only one motive $M_E$, leading to the 
 first cohomology group
  \begin{equation}
  H(M_E) ~=~ H^1(E).
  \end{equation}
  The conjectures of Taniyama, Shimura, and in particular Weil made it
  possible to determine for any given elliptic curve $E$ the modular form of the
  $L-$series $L(M_E,s)$ by computing the conductor $N$ of the
  elliptic curve $E$ and comparing the result with the basis
  vectors in the finite dimensional space $S_2(\Gamma_0(N))$.

For higher dimensional Calabi-Yau manifolds (and more general
spaces) a universal construction was given in \cite{rs08},
generalizing to all Calabi-Yau varieties the special case of weighted 
hypersurfaces described in \cite{rs06}. This construction is
based on the existence of the intermediate cohomology group, in
particular the existence of the holomorphic $(n,0)-$form, usually
called $\Omega \in H^{n,0}(X)$, for a Calabi-Yau $n-$fold $X$. For
this reason the motive derived from this form is called the
$\Omega-$motive \cite{rs06,rs08}. The generalization of the
$\Omega-$motive to a more general class of special Fano varieties
  \cite{rs92, cdp93, rs94} leads to a "twisted" cohomology group
$H^{n-w,w}(X)$ \cite{kls08}. More precisely, the construction of this 
motive is based on a Galois
group ${\rm Gal}(K_X/{\mathbb Q})$ of a number field $K_X$ that is
derived from the arithmetic structure of the manifold.

 For modular $\Omega-$motives of Calabi-Yau manifolds
  the relation (\ref{weight}) between the
 weight of the modular form and the weight of the motive leads to
 a direct relation between the weight of the modular form $f_\Omega$
 associated to $M_\Omega$ and the dimension of the variety as
 \begin{equation}
  w(f_\Omega) ~=~ {\rm dim}_{\mathbb C}X + 1.
 \end{equation}
 For CY threefolds the relevant forms are therefore of weight
 four and some level $N$,
  $f_\Omega \in S_4(\Gamma_0(N))$, where $N$ is
 determined by the structure of the motive $M_\Omega$ in a way
 that is not understood at present.

\subsection{A speculative framework for a proof}

It is not clear a priori what the correct framework should be in
which one might attempt to establish a proof of the relation
conjectured to exist between the modular forms on the worldsheet
$\Sigma$ on the one side, and modular, or automorphic, motives of
spacetime on the other. It is nevertheless tempting to identify
the structures that appear in the examples considered so far, and
to speculate what might be some of the concepts that should be key
ingredients of a proof. First, on the geometric side of the conjectured 
 relation there should exist a set of automorphic forms that are associated to
 motives that arise in spaces that are either of
 Calabi-Yau type or of special Fano type. Denote the
set of all motives associated to Calabi-Yau and special Fano type
varieties by ${\cal M}_{\rm CYF}$, and denote the set of all
automorphic forms of all motives by ${\cal M}_{\rm CYF}$ by
${\cal A}({\cal M}_{\rm CYF})$. On the string theoretic
 side one might envision a second set
  ${\cal A}({\rm CFT}_{{\cal N}=2})$ consisting of automorphic forms that arise
 from conformal field theories that live on the string worldsheet $\Sigma$.
 The hoped-for relation would then take the form of an
 identification of these two rather large sets
 \begin{equation}
 {\cal A}({\rm CFT}_{{\cal N}=2}) ~=~ {\cal A}({\cal M}_{\rm CYF}).
 \label{goal}\end{equation}

The above outline is too simplistic and has too little structure
to be of ultimate use. It is natural to expect that a proof of the
relation (\ref{goal}) is best approached in terms of structures
that allow to build complicated objects from simpler ones, which
means that one should endow the sets considered above with
categorical notions. On the geometric side this is possible at
present for the case of smooth manifolds. In this case there
exists a category ${\cal M}$ of so-called pure motives. This
category admits a tensor construction, hence it can be viewed as
providing the building blocks of smooth projective
manifolds. The existence of phase transition between Calabi-Yau
manifolds \cite{cdls88}, leading to the idea of a connected
universal moduli space of string compactifications
\cite{cgh89,cgh90}, necessitates the consideration of the more
general concept of mixed motives, as shown in \cite{kls10}. The
notion of a category of mixed motives has proven to be problematic
and has not been accomplished yet in a satisfactory way. On the string theoretic side
the idea would be to consider a category of conformal field
theories with ${\cal N}=2$ supersymmetry, as well as their
associated automorphic forms.  At present the notion of a category
of conformal field theories has also not been formalized.

\subsection{An elliptic example}

This subsection illustrates the construction of the extra
dimensions from string worldsheet modular forms in the simplest
possible example, that of a compactification of string theory to
eight dimension on a two-torus. More details can be found in
\cite{su02}. Suppose the worldsheet theory is given by the Gepner
model based on the tensor product $1^{\otimes 3}$ of three 
 ${\cal N}=2$ superconformal minimal
models, each with conformal level $k=1$ and central charge $c(k) =
3k/(k+2)=1$. The first problem that arises is that in general 
such minimal
models lead to many modular forms, and it is a priori not obvious
which modular form one should consider. It turns out that the key
structure that is of importance are the Hecke indefinite modular
forms, given in terms of the parafermionic partition functions
given by the Kac-Petersson string functions $c^k_{\ell,m}(\tau)$
\cite{kp84}. More precisely, the correct form to consider is the
weight one forms
 \begin{equation}
  \Theta^k_{\ell,m}(\tau) ~=~ \eta^3(\tau) c^k_{\ell,m}(\tau),
 \end{equation}
 where $\eta(\tau)$ is the Dedekind eta function.

The example $1^{\otimes 3}$ is a make-or-break case because  for 
conformal level $k=1$ the model has only a single independent Hecke indefinite
 modular form $\Theta^1_{1,1}(\tau)$, hence there is no 
 room to maneuver $-$ this form either leads to the correct elliptic curve, or the 
whole set-up fails.
 Elliptic curves $E_N$ of conductor $N$ are known to be modular in the sense
 that the inverse Mellin
 transform of their $L-$function are modular forms of weight two
 with respect to the Hecke congruence subgroup of level $N$,
 i.e. $\Gamma_0(N) \subset {\rm SL}(2,{\mathbb Z})$
 \cite{w95, tw95, bcdt01}. In
 concrete cases this can always be checked by a finite computation
 without the Galois machinery of Wiles et al. The general relation
 between the dimension of a Calabi-Yau variety and the weight of the modular
 form tells us that for elliptic curves the motivic form has to
 have weight two. This means that the motivic form should be a
 product $\Theta^1_{1,1}(a\tau) \Theta^1_{1,1}(b\tau)$ for some
 integers $a,b\in {\mathbb N}$. These integers can be determined
 by a variety of criteria described in detail in \cite{rs05},
 leading to a weight two form of level $N=27$
 \begin{equation}
 f_2(\tau) ~=~ \Theta^1_{1,1}(3\tau) \Theta^1_{1,1}(9\tau) \in
 S_2(\Gamma_0(27)).
 \end{equation}
 From this modular form the elliptic curve of conductor 27 can be
 determined to be the cubic Fermat curve embedded in projective
 space ${\mathbb P}_2$, consistent with previous conjectures.

 \subsection{Higher dimensional diagonal varieties and their
 families}

Higher dimensional extensions of the example above have been described
 both for the case of diagonal Calabi-Yau hypersurfaces
 \cite{rs06, rs08}, diagonal weighted hypersurfaces
 of special Fano type \cite{kls08}, as well as for families of Calabi-Yau varieties \cite{kls10}. 
 The diagonal varieties considered in
 these references are related to Gepner models, constructed via
 tensor products of central charge $c=3D$, where $D={\rm
 dim}_{\mathbb C}X$, of ${\cal N}=2$ supersymmetric minimal
 models. This class of models \cite{g88} has been constructed
 explicitly, and their cohomological spectra have been computed in
 \cite{ls89, fkss90}.

 Diagonal varieties are special because their underlying conformal
 field theory is rational. These spectra of these CFTs contain
 marginal operators, and it is of importance to consider
 deformations of the diagonal points. Conformal field
 theoretically, such deformations have proven difficult, and much
 less is known about them. In \cite{kls10} a first exploration was
 initiated of modular motives that arise from singular Calabi-Yau
 varieties that appear in families of deformations of weighted
 hypersurfaces. This work uncovered some unexpected features of
 modular phases that involve non-conifold type singularities.
 First, depending on the degree of the singularity, degenerate
 $\Omega-$motives can be modular, even though naively they appear
 to be of high rank (hence in the smooth case would lead to
 automorphic forms, rather than modular forms). Second, the
 modular forms that can be identified from the $L-$functions of these
 degenerate $\Omega-$motives can sometimes be shown to arise from
 modular forms that appear in smooth weighted Fermat hypersurfaces. 
 This means that
 highly degenerate deformations of Gepner models can lead to the
 same modular forms that arise in the Gepner models themselves.

 The phenomena observed in \cite{kls10} can be interpreted as an
 indication that highly
 singular Calabi-Yau varieties that appear as intermediate phases
 in transitions between different Calabi-Yau manifolds define
 consistent string configurations.

\subsection{Arithmetic mirror symmetry}

The construction of the compact geometry directly from the
worldsheet theory leads to a number of new questions in the
context of mirror symmetry. The discovery of mirror symmetry in
the context of weighted Calabi-Yau hypersurfaces \cite{cls90}
 can be combined with the quotient mirror construction of Greene and
 Plesser \cite{gp90} to formulate a mirror map directly at the
 level of Calabi-Yau varieties \cite{ls90}. Since the quotient
 construction itself leads to an isomorphic conformal field theory
 it is clear that the mirror theory contains the same modular
 forms as the original theory, hence one might expect that certain
 arithmetic aspects of the mirror manifold are identical to that
of the original theory. This notion is problematic because the
cohomology groups of mirror pairs are completely different. While
this problem has not been resolved (see \cite{cdov00}), there are
some simple tests that can be performed. The simplest possible
case is to consider elliptic compactifications, i.e. Gepner tensor
models with total central charge $c=3$, of which there are only
three. The mirror construction of \cite{ls90} leads to a mirror
pairing, and one can ask whether these mirror pairs have the same
$L-$function. The answer to this is affirmative \cite{rs07},
leading to a first arithmetic test of the geometric mirror
construction of \cite{ls90}.

A key problem in mirror symmetry is to find the appropriate class
of varieties that allows to construct mirrors of rigid Calabi-Yau
manifolds. One class of varieties that has been suggested as such
a framework is defined by a special class of Fano varieties
\cite{rs92,cdp93,rs94}. These varieties are of higher dimension
than their Calabi-Yau counterparts, but contain motives that
encode information about mirrors. It was shown in \cite{kls08}
that the $L-$series of the Fano varieties associated to two
particular Gepner models are modular, and that the corresponding
modular forms agree with the modular forms previously computed
\cite{ch05} for two rigid Calabi-Yau threefolds whose cohomology
is mirror symmetric to that of the Fano varieties.

\section{Automorphic black holes as probes of extra dimensions}

The remainder of this review is dedicated to the description of a
second recent program to apply ideas of the arithmetic
Langlands program in physics. The problem posed here is what
information could be extracted from black holes if we were able to
experiment with them in the laboratory. More precisely, if the
entropy of black holes is described by automorphic forms, as
outlined below, then one might expect that such 
automorphic black holes encode information about the geometry 
of the extra dimensions in
string theory. This raises the question of how one can identify
the motives of the variety which support the automorphic forms
that appear in the entropy results. This problem was addressed in
\cite{cs11}, on which the discussion of this section is based.

\subsection{Automorphic black holes}

Over the past 15 years impressive progress has been made toward
the resolution of a problem that is almost 40 years old - the
microscopic understanding of the entropy of black holes. It has
been proven useful to focus on black holes with extended
supersymmetries because this leads to black holes that are simple,
but not too simple. It was shown in particular for certain types
of black holes in ${\cal N}=4$ supersymmetric theories that their
entropy is encoded in the Fourier coefficients of
Siegel modular forms, automorphic forms that form one of the
simplest generalizations of classical modular forms of one
variable with respect to congruence subgroups of the full modular
group ${\rm SL}(2,{\mathbb Z})$.

The general conceptual framework of automorphic entropy functions
has not been formalized yet. A formulation that generalizes
the existing examples can be outlined as follows. Suppose
we have a theory which contains scalar fields parametrized by a
homogeneous space $\prod_i (G_i/H_i)$, where the $G_i$ are Lie
groups. Associated to these scalar fields are electric and
magnetic charge vectors $Q=(Q_e,Q_m)$, taking values in a lattice
$\Lambda$ whose rank is determined by the groups $G_i$.

Assume now that the theory in question has a T-duality group
$\prod_i D_i({\mathbb Z})$, where $D_i({\mathbb Z}) \subset
G_i({\mathbb Z})$ denotes the Lie groups considered over the
rational integers ${\mathbb Z}$. Suppose further that the charge
vector $Q$ leads to norms $||Q||_i, ~i=1,...,r$ that are invariant
under the T-duality group. Choose conjugate to these invariant
charge norms complex chemical potentials
 \begin{equation}
 (\tau_i, ||Q||_i),~~ i=1,...,r,
 \end{equation}
which generalize the upper half plane of the bosonic string. On
this generalized upper half plane ${\cal H}_r$ formed by the
variables $\tau_i$ one can consider automorphic forms
$\Phi(\tau_i)$, and the idea is that with an appropriate integral
structure ${\mathbb Z} \ni k_i \sim ||Q||_i, i=1,...,r$
 associated to the charge norms, the Fourier expansion of these
 automorphic forms given by
 \begin{equation}
  \Phi(\tau_i) ~=~ \sum_{k_n \in {\mathbb Z}} g(k_1,...,k_r)
                    q_{1}^{k_{1}} \cdots q_{r}^{k_{r}} ,
 \end{equation}
 in terms of $q_k = e^{2\pi i \tau_k}$,
 determines the automorphic entropy via the coefficients of the
 expansion of the automorphic partition function
 \begin{equation}
  Z ~=~ \frac{1}{{\widetilde \Phi}} ~=~ \sum_{k_n}
  d(k_1,...,k_r) q_{1}^{k_{1}} \cdots q_{r}^{k_{r}},
 \end{equation}
 as
 \begin{equation}
  S_{\rm mic}(Q) ~\sim ~ \ln d(Q),
 \end{equation}
 where
 \begin{equation}
 d(Q) := d(||Q||_1,...,||Q||_r).
 \end{equation}
 Here ${\widetilde \Phi}$ denotes a modification of the Siegel form
 $\Phi$ that is determined by the divisor structure of $\Phi$.

\subsection{Siegel modular black holes in ${\cal N}=4$ theories}

The above automorphic-entropy-outline describes the behavior of
the entropy of black holes in certain ${\cal N}=4$
compactifications obtained by considering ${\mathbb
Z}_N-$quotients of the heterotic toroidal compactification
 ${\rm Het}(T^6)$, a small class of models first considered by
Chaudhuri-Hockney-Lykken models
 \cite{chl95}. Specifically, it was shown in \cite{dvv96,js05,gk09}
 that for these CHL$_N$ models the microscopic entropy of extreme
 Reissner-Nordstrom type black holes is
described by Siegel modular forms $\Phi^N \in
S_w(\Gamma^{(2)}_0(N))$, where the weight $w$ is determined by the
order $N$ of the quotient group. In this case the dyonic charges
$Q=(Q_e,Q_m)$ form three integral norms invariant under the
T-duality group ${\rm SO}(6,n_v)$, where $n_v$ depends on the
group order $N$. These norms are usually denoted by
 \begin{equation}
 ||Q||_1 = \frac{1}{2}Q_e\cdot Q_e,~~~
 ||Q||_2 = \frac{1}{2}Q_m\cdot Q_m,~~~
 ||Q||_3 = Q_e\cdot Q_m,
\end{equation}
 where the inner product $\cdot$ is defined with respect to the
 defining metric of the non-compact duality group $G({\mathbb R})$.
 The conjugate complex variables $\tau_1,\tau_2,\tau_3$ form
 the Siegel upper half plane ${\cal H}_2$
 \begin{equation}
 Z ~=~ \left(\matrix{
              \tau_1  &\tau_3 \cr \tau_3 &\tau_2\cr}\right)  \in {\cal H}_2.
 \end{equation}
 The automorphic groups are Hecke type congruence subgroups
 $\Gamma_0^{(2)}(N) \subset {\rm Sp}(4,{\mathbb Z})$, hence the associated
 forms are Siegel modular forms of genus two.

The key feature of the Siegel modular forms that appear in the
context of CHL$_N$ black hole entropy is that they are not of
general type, but belong to the Maa\ss~Spezialschar, i.e. they are
obtained via a combination of the Skoruppa lift \cite{nps92} from
classical modular forms to Jacobi forms, and the Maa\ss~lift \cite{m79} 
 from
Jacobi forms to Siegel modular forms
 \begin{equation}
  f(\tau) \in S_{w+2} ~\stackrel{\rm SL}{\longrightarrow} ~\varphi_{w,1}(\tau,\rho) \in J_w
                      ~\stackrel{\rm ML}{\longrightarrow} ~\Phi_w(\tau,\sigma,\rho) \in S_w,
 \end{equation}
 where $\tau=\tau_1, \sigma=\tau_2, \rho=\tau_3$. The classical
 modular form $f(\tau)$ whose Maa\ss-Skoruppa lift is the Siegel modular form
 $\Phi_w = {\rm MS}(f)$ is called the Maa\ss-Skoruppa root.

\subsection{Automorphic motives}

Given that the entropy of black holes is described by automorphic
forms, one can ask whether the spacetime structure of the
compactification manifolds leads to motives rich enough to support
these automorphic forms. It is not expected that general
automorphic forms are of motivic origin, however algebraic
automorphic forms are conjectured to be supported by motives.
 Background material for Siegel forms and motives can be found in
 \cite{y01, p11}.

In the special case of Siegel forms of genus two modular forms
that appear in the context of CHL$_N$ black holes the conjectures
concerning the motivic origin indicate that the compactification
manifold cannot provide directly the appropriate motivic cycle
structure. The easiest way to see this is as follows \cite{cs11}.
Suppose $M_\Phi$ is a motive whose $L-$function $L(M_\Phi,s)$
agrees with the spinor $L-$function $L_{\rm sp}(\Phi,s)$ associated
to a Siegel modular form $\Phi$ of arbitrary genus $g$ and weight
$w$
 \begin{equation}
  L(M_\Phi,s) ~=~ L_{\rm sp}(\Phi,s).
 \end{equation}
 The weight ${\rm wt}(M_\Phi)$ of such genus $g$ spinor motives follows from the (conjectured)
 functional equation of the $L-$function as
 \begin{equation}
  {\rm wt}(M_\Phi) ~=~ gw - \frac{g}{2}(g+1).
 \end{equation}
 For the special case of genus 2 spinor motives the Hodge
 structure takes the form
 \begin{equation}
  H(M_\Phi) ~=~ H^{2w-3,0} \oplus H^{w-1,w-2} \oplus
                H^{w-2,w-1} \oplus H^{0,2w-3}.
 \label{siegel-hodge}\end{equation}
 It should be noted that this Hodge structure only applies to pure
 motives. In the case of mixed motives it is possible, for example, 
 that rank 4 motives can give rise to classical modular forms  \cite{kls10}.

 While the Hodge (\ref{siegel-hodge}) type of $M_\Phi$ is that of a Calabi-Yau variety,
 the precise structure is only correct for modular forms of weight
 three. It turns out that for the class of CHL$_N$ models the
 weights of the Siegel modular forms take values in a much wider
 range $w \in  [1,10]$.
 If follows that for most CHL$_N$ models the Siegel modular form
 will take the wrong value to be induced directly by motives in
 the way usually envisioned in the conjectures of arithmetic geometry.

The same is the case for the classical Maa\ss-Skoruppa roots,
whose weights are given by $(w+2)$. The motivic support $M_f$ for
such modular forms $f$ is of the form
 \begin{equation}
 H(M_f) ~=~ H^{w-1,0} \oplus H^{0,w-1},
 \end{equation}
 hence the only modular forms that can fit into heterotic
 compactifications have weight two, three, or four.

\subsection{Lifts of weight two forms}

The key to the identification of the motivic origin of the CHL$_N$
black hole entropy turns out to be an additional lift construction
that interprets the Maa\ss-Skoruppa roots of weight $(w+2)$ in
terms of modular forms of weight two for all $N$, and hence in
terms of elliptic curves \cite{cs11}. These Maa\ss-Skoruppa roots
decompose into two distinct classes of forms, one class admitting
complex multiplication, the second class not. For this reason it
is not surprising that the lifts of weight two modular forms to
the CHL$_N$ Maa\ss-Skoruppa roots involve two different
constructions, depending on the type of the higher weight form.
For the forms without complex multiplication the lift
interpretation of the MS root $f_{w+2}$  in terms of the weight
two form $f_2 \in S_2$ can be written as
 \begin{equation}
  f_{w+2}(q) ~=~ f_2(q^{1/m})^m, ~~~~{\rm with}~~m =
  \frac{1}{2}\left\lceil \frac{24}{N+1}\right\rceil.
  \end{equation}
 The lift for the class of Maa\ss-Skoruppa roots with complex
 multiplication derives from the existence of algebraic Hecke
 characters whose $L-$functions are the inverse Mellin transform
 of the MS roots. More details can be found in ref. \cite{cs11}.

 The interpretation of the Maa\ss-Skoruppa roots in terms of weight 2 modular
 forms $f_2^{\tilde N}$ via these two additional lifts for CM and non-CM forms
 shows that the motivic orgin of the Siegel modular
 entropy of CHL$_N$ models is to be found in elliptic curves. This follows
 from the fact that for all CHL$_N$ models the geometric structure that
 supports the weight 2 forms is that of elliptic curves $E_{\tilde N}$,
 whose conductor ${\tilde N}$ varies with the order $N$ of the quotient
 group ${\mathbb Z}_N$. More precisely, the $L-$functions associated to both of
 these objects agree
  \begin{equation}
   L(f_2^{\tilde N},s) ~=~ L(E_{\tilde N},s).
  \end{equation}
 Abstractly, this follows from the proof of the
 Shimuar-Taniyama-Weil conjecture \cite{w95} and \cite{bcdt01},
 but no such heavy machinery is necessary for the concrete cases
 based on the CHL$_N$ models, where the elliptic curves can be
 determined explicitly for each $N$.
  This shows that the motivic origin of the
 Siegel black hole entropy considered
 in \cite{dvv96,js05,gk09} can be reduced to that of
 lower-dimensional cycles, as opposed to the three-dimensional
 nature of the compactification manifold $X_N=T^6/{\mathbb Z}_N$ in the
 heterotic frame, or $(K3\times T^2)/{\mathbb Z}_N$ in the type IIA
 frame.

\vskip .2truein

\noindent
 {\bf Max} 

During the course of the work described here I often had reason to
think of one of Max's proverbs, which I learned from him. This was
in particular the case during the early stages of the program to
use motivic modular forms to formulate a new approach to the
problem of an emergent spacetime.  The relation between
CFT-theoretic and motivic modular forms that I was looking for
initially remained hidden and it seemed for some time that it actually 
might not exist. The main problem was that it is a priori
not clear which, if any, conformal field theoretic modular forms
on the worldsheet might be useful as building blocks of the
motivic $L-$series of the compact dimensions. This was cause for
some aggravation, and while thinking about this I often recalled
one of Max's favorite comments, who in such times reminded himself
and his friends of the proverb
 "M\"uhsam ern\"ahrt sich das Eichh\"ornchen".
 It is an interesting open problem to find a translation
  that properly conveys the meaning of this important insight into
  human nature.

\vskip .2truein
 \noindent
{\bf Acknowledgement} 

 This contribution to Max's Memorial Volume was written in part
 at the Max Planck Institute for Physics. It is a pleasure to thank
 Dieter L\"ust for making my visit possible, and the MPI for hospitality
 and support.
 It is a pleasure to thank Ilka Brunner, Stefan Hohenegger, Monika Lynker
 and Stephan Stieberger for discussion.
 The work reviewed here was supported in part by the National Science Foundation
 under grant No. PHY 0969875.

\vskip .2truein

\bibliographystyle{ws-rv-van}

\end{document}